\documentclass[aps,prb,twocolumn,superscriptaddress]{revtex4} 

\usepackage{prettyref}
\usepackage{textcomp}
\usepackage{amsmath,amssymb}
\usepackage{graphicx}

\begin{document}

\title{Metamaterial tuning by manipulation of near-field interaction}

\author{David A. Powell}
\email{david.a.powell@anu.edu.au}
\affiliation{Nonlinear Physics Centre, Research School of Physics and Engineering,\\
 Australian National University, Canberra ACT 0200, Australia }

\author{Mikhail Lapine}
\affiliation{Dept. Electronics and Electromagnetics, Faculty of Physics, University
of Seville, Avda. Reina Mercedes s/n, 41015 Seville, Spain}

\affiliation{Nonlinear Physics Centre, Research School of Physics and Engineering,\\
Australian National University, Canberra ACT 0200, Australia }

\author{Maxim V. Gorkunov}
\affiliation{A. V. Shubnikov Institute of Crystallography, Russian
Academy of Sciences, Lenin Ave. 59, 119333 Moscow, Russia}

\author{Ilya V. Shadrivov}
\affiliation{Nonlinear Physics Centre, Research School of Physics and Engineering,\\
 Australian National University, Canberra ACT 0200, Australia }

\author{Yuri S. Kivshar}
\affiliation{Nonlinear Physics Centre, Research School of Physics and Engineering,\\
 Australian National University, Canberra ACT 0200, Australia }

\begin{abstract}
We analyze the near-field interaction between the resonant sub-wavelength elements of a metamaterial,
and present a method to calculate the electric and magnetic interaction coefficients.
We show that by adjusting the relative configuration of the neighboring split ring resonators
it becomes possible to manipulate this near-field interaction, and thus tune the response of 
metamaterials. We use the results of this analysis to explain the experimentally observed 
tuning of microwave metamaterials.
\end{abstract}

\maketitle

\section{Introduction}

Metamaterials, which are typically regular arrays of
sub-wavelength resonant particles, offer us a new degree of
freedom in controlling the electromagnetic response of matter.
Thus we are no longer completely constrained by the properties of
existing materials, but can tailor the response in an almost
arbitrary fashion, for example achieving very high \cite{Silveirinha2008c}, very low
\cite{Ziolkowski2004}, and negative\cite{Pendry1999} values of refractive index,
permittivity and/or permeability.
Because of the inherently strong dispersion of
resonant metamaterials, they must be modified in order to operate in a
different frequency band. Therefore,
there is a significant push to have a further degree of control
over these materials --- tunability of their response.

Fortunately, the engineered nature of metamaterials allows their
properties to be controlled externally, either by dynamically
modifying their structure, or by adding some nonlinear inclusion
and controlling with external fields~\cite{Gorkunov2004prb}.
Examples of the latter approach include the introduction of
varactor diodes~\cite{Shadrivov2006a}, ferroelectrics~\cite{Hand2008} and photo-conductive
semi-conductors~\cite{Degiron2007}. On the other hand,
even without resorting to such exotic (and often lossy)
constituents, there is a great deal of freedom to manipulate the
structure itself, and this is the approach we take here. We
consider specifically the split ring resonator (SRR) as one of the
most important metamaterial elements, noting that whilst the
fine details of near-field interaction are structurally specific,
our approach can be applied to a wide variety of structures.

An analytical model for the magnetic response of a sub-wavelength array
of identically oriented wire loops loaded with
a capacitance~\cite{Gorkunov2002} takes
into account the mutual interaction of all the elements in the
lattice, which is essential for deriving the effective
permeability correctly. Although that analysis is limited to the
quasi-static case accounting only for magnetic near-field
interactions, it is crucial for revealing the consequence of
lattice changes. These tend to be overlooked by
otherwise rigorous approaches which include spatial dispersion but
develop Lorentz local field approaches,
based on nearest-neighbor interaction~\cite{Baena2008}
or point-dipole approximation~\cite{Simovski2008mm}.

\begin{figure}
\includegraphics[width=0.8\columnwidth]{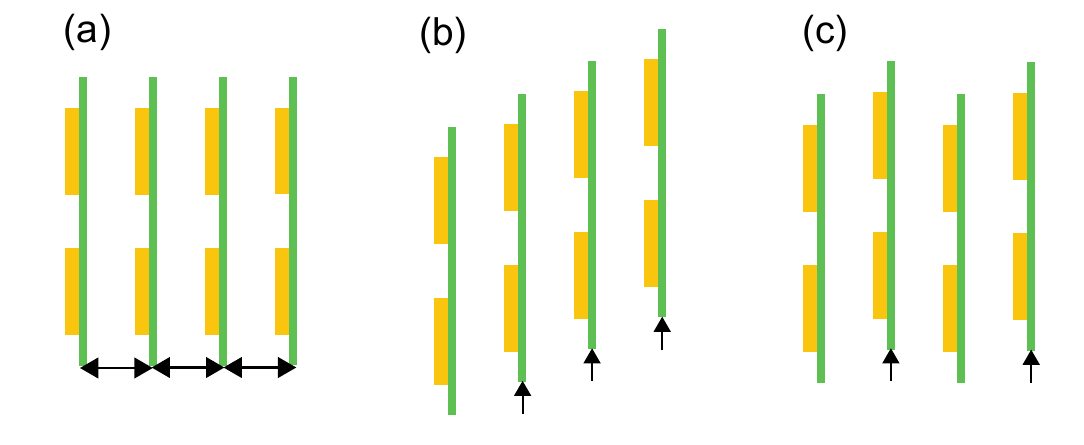}\caption{(Color online)
Several approaches to modify the lattice for metamaterial tunability:
(a) a change of the lattice constant~\cite{Shadrivov2007a}, (b)
a continuous shift of the layers~\cite{Lapine2009}; and
(c) a super-lattice of alternating shifts of layers~\cite{Lapine2009}.
\label{fig:lattice-tuning-schemes}}
\end{figure}

In particular, it was pointed out in
Ref.~\onlinecite{Gorkunov2002} that the resonant frequency of the
metamaterial permeability can be altered by varying the lattice
constants without changing the structural units. This scheme is
illustrated in Fig.~\ref{fig:lattice-tuning-schemes}(a), and has
been verified by experiments in Ref.~\onlinecite{Shadrivov2007a}.
However, a practical consequence of this change in lattice
constant is that the sample size also changes correspondingly.
More recently, an alternative approach was suggested in
Ref.~\onlinecite{Lapine2009}: introducing a shift between layers
in order to create a monoclinic lattice, with the shift increasing
linearly between layers, as shown in
Fig.~\ref{fig:lattice-tuning-schemes}(b). This configuration keeps
the density of elements within the metamaterial constant, and can
tune the coupling between neighboring particles to modify the
response of the complete metamaterial. However, for finite size
samples, this shift inevitably results in a significant change in
sample shape. Thus, for practical purposes, we have proposed a
super-lattice type of geometry, whereby only every second layer is
shifted by the same amount, as shown in
Fig.~\ref{fig:lattice-tuning-schemes}(c). This tuning scheme
proved to be robust, and allows significant manipulation of the
resonant frequency with only a small change in the sample
geometry~\cite{Lapine2009}. Therefore, the sample geometry and its
effective properties can be engineered {\em almost independently} to
achieve the desired manipulation of electromagnetic waves.

However, as we demonstrate below, this structural tuning of
metamaterials depends very strongly on the nature of the near-field interactions. 
Since metamaterial elements such as split ring
resonators are usually not highly symmetric, \emph{the relative
orientation of particles within the lattice is of key importance}.
This effect is not described by existing circuit theory models,
and can give rise to some surprising experimental results, which
we present here.

In order to understand the coupling mechanisms and how they are
affected by the lattice shift, it is useful to start with the
simplest geometry --- a pair of split ring resonators. Several authors have
conducted numerical and experimental investigations of coupling
between metamaterial elements for different types of elements and
relative orientation between them,
in microwave~\cite{Garcia-Garcia2005,Gay-Balmaz2002}
and optical frequency ranges.(see the overview in Ref.~\onlinecite{Liu2009b} and references therein,
as well as Refs.~\onlinecite{Li2009a,Kante2010,Zhou2009a,Decker2009,Singh2009a,Liu2008}.)
A detailed study has previously been undertaken on tailoring the geometric arrangement
of a pair of coupled one-dimensional SRR arrays to engineer the dispersion
curves of magneto-inductive waves \cite{Sydoruk2006}.

As we have shown recently~\cite{Lapine2009}, a rough qualitative
understanding of the structural tuning can be achieved by using
circuit theory with purely inductive coupling between SRRs.
However, this approximation fails to provide a quantitative
understanding of the metamaterial tuning in question and, most
importantly, does not explain the observed strong influence of the
relative SRR orientation.
Therefore, below we develop {\em a new model} based on the calculation of 
the fundamental mode of a single resonator. The knowledge of
the current and charge distributions within the mode allows
calculating the coupling constants of a pair of split rings. These
constants are then used to explain the experimentally observed
tuning response of our metamaterial samples.

In \prettyref{sec:theory}, we develop our approach to calculate the metamaterial coupling,
including a discussion of the limitations of other purely analytical methods.
In \prettyref{sec:pair}, we apply our approach to the study of
interaction between a pair of split ring resonators which are shifted 
laterally relative to each other, and explain quantitatively how the shift affects
the position of the fundamental resonance. In \prettyref{sec:slab}, we
apply these results to a bulk metamaterial and identify the
mechanisms at work in the experimental tuning of a metamaterial
slab in a waveguide. Finally, \prettyref{sec:concl} concludes the
paper with further discussions and outlook.

\section{Near-field interaction in metamaterials\label{sec:theory}}

Considering a single SRR, it is known\cite{Garcia-Garcia2005} 
that it possesses a discrete set of
eigenmodes (standing waves) with corresponding eigenfrequencies.
In an arbitrarily excited SRR, the currents and charges can be
represented as a superposition of the eigenmodes. The fundamental mode with
the lowest frequency is relevant for the
magnetic resonance in SRRs. On the frequency scale this mode is well
isolated from the higher-order modes, and we can restrict
ourselves to the single-mode approximation neglecting the
excitation of higher modes.

The time dependent charge density $\rho$ and current density
$\mathbf{J}$ in a resonant element with excited fundamental mode
can be written in the most general form of a standing wave:
\begin{equation}\label{eq:Qwave}
\rho(\mathbf{x},t)=Q(t)q(\mathbf{x}),
\end{equation}
\begin{equation}\label{eq:Iwave}
\mathbf{J}(\mathbf{x},t)=I(t)\mathbf{j}(\mathbf{x}),
\end{equation}
where $q$ and $\bf J$ describe the charge and current distribution
in space. In SRRs, the variation of the current distribution across the
width of the conductive track could be neglected, however
for generality our approach takes into account the complete
3-dimensional surface-current distribution.

\begin{figure}[t]
\includegraphics[width=\columnwidth]{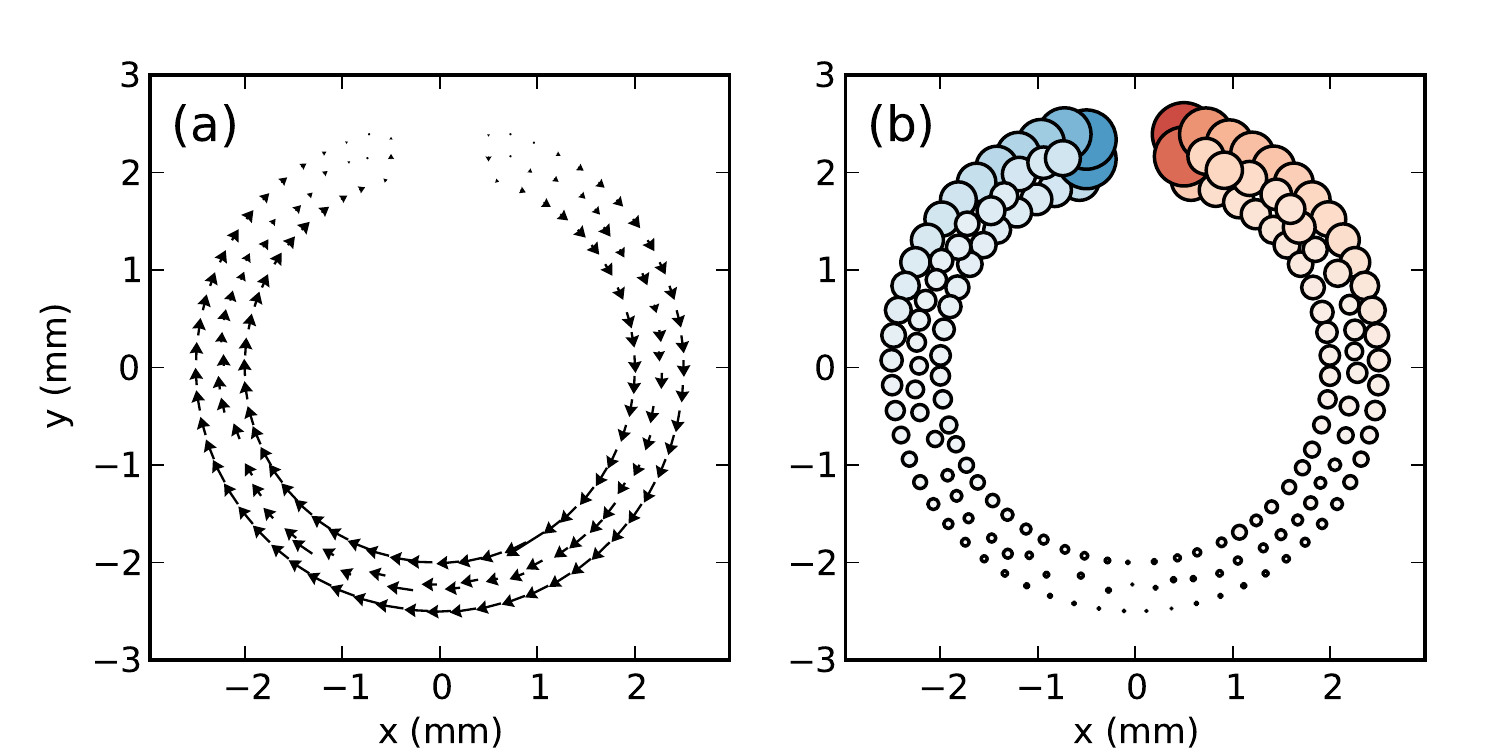}
\caption{Numerically calculated (a) current and (b) charge distribution across an SRR at resonance\label{fig:current_charge}}
\end{figure}

To satisfy the conservation of charge
\begin{equation}
\nabla \cdot \mathbf{J}=-\frac{\partial \rho(\mathbf{x},t)}{\partial t}
\end{equation}
we imply that
\begin{equation}\label{eq:QI}
I(t)=\dot{Q}(t),
\end{equation}
\begin{equation}\label{eq:qj}
\nabla \cdot \mathbf{j}(\mathbf{x})=-q(\mathbf{x}).
\end{equation}
Thus if the the current is known, it is easy to find the charge
distribution and vice-versa. The mode profile obtained numerically
for our SRR geometry is shown in Fig.~\ref{fig:current_charge}.
The current $\mathbf{j}$ is symmetric and reaches its maximum at
the point opposite to the gap. In accordance with
Eq.~\eqref{eq:qj}, the charge distribution $q(\mathbf{x})$ is
antisymmetric and goes through zero where $\mathbf{j}(\mathbf{x})$
is maximal. We see that $q(\mathbf{x})$ reaches its maximum magnitude near the gap.

In the single mode approximation, the dynamics of the SRR can be
fully described by the time-dependent amplitude $Q(t)$, and we may
write the SRR Lagrangian as a sum of terms quadratic in $Q$ and
$\dot{Q}$:
\begin{equation}
\mathcal{L}=A\dot{Q}^2-BQ^2,
\end{equation}
where $A$ and $B$ are constants which will be discussed below.
Accordingly, the SRR energy reads:
\begin{equation}
E=\dot{Q}\frac{\partial \mathcal{L}}{\partial\dot{Q}}-\mathcal{L}= AI^2+BQ^2,
\end{equation}
and is nicely separated into inductive (magnetic) and capacitive
(electric) parts. Clearly, for a passive SRR we require $A \geqslant 0$,
$B \geqslant 0$.

The Lagrangian equation of motion
\begin{equation}
\frac{\mathrm d}{\mathrm d t} \frac{\partial \mathcal{L}}{\partial\dot{Q}} =
\frac{\partial \mathcal{L}}{\partial Q}
\end{equation}
yields that the dynamics of a single SRR is described by the
oscillator equation for the charge amplitude:
\begin{equation}
\ddot{Q}(t)+\omega_0^2Q(t)=0,
\end{equation}
and the fundamental mode resonance occurs at the frequency
$\omega_0=\sqrt{B/A}$.

Note that in contrast to the known modifications of the Lagrangian
formalism for metamaterials (see e.g.\ Ref.~\onlinecite{Liu2006c}), here
we do not rely on an equivalent circuit model. Although one might
identify the parameters $2A$ and $1/2B$ as effective inductance
and capacitance respectively, below we evaluate them explicitly
from the known fundamental mode shape. In fact, the strongly
inhomogeneous mode profile (see Fig.~\ref{fig:current_charge})
suggests that it is unlikely that the correct values of the
parameters would agree with those calculated from a circuit
analysis. Additionally, to find the resonant frequency, we do not
need to calculate $A$ and $B$ explicitly, only their ratio.

For the purposes of our analysis, it is sufficient to consider the
case of a pair of SRRs, and it will subsequently be shown that
this explains all of the important features observed in our
experiments with a bulk metamaterial. In this case, the Lagrangian
can be written as a sum of the single SRR Lagrangians and coupling
terms, which we also write as quadratic in currents and charges:
\begin{equation}\label{eq:L12}
\mathcal{L}=A(\dot{Q}_1^2+\dot{Q}_2^2+2\alpha
\dot{Q}_1\dot{Q}_2)-B(Q_1^2+Q_2^2+2\beta Q_1Q_2).
\end{equation}
The parameters $\alpha$ and $\beta$ are the dimensionless
constants of magnetic and electric near-field interaction
respectively.

The corresponding Lagrangian equations of motion yield the system
of ODEs for the time-dependent amplitudes $Q_{1,2}$:
\begin{eqnarray}\label{eq:Q1Q2}
\ddot{Q}_1+\omega_0^2 Q_1=-\alpha \ddot{Q}_2-\beta \omega_0^2 Q_2,\\
\ddot{Q}_2+\omega_0^2 Q_2=-\alpha \ddot{Q}_1-\beta \omega_0^2 Q_1,
\end{eqnarray}
Solving these equations one finds that a pair of resonators
exhibits two resonances: symmetric and antisymmetric. For the
symmetric resonance, $Q_1=Q_2$, which yields the resonant
frequency
\begin{equation}\label{eq:omega_s}
\omega_{S}=\omega_0\sqrt{\frac{1+\beta}{1+\alpha}},
\end{equation}
while the antisymmetric mode with $Q_1=-Q_2$ has the frequency
\begin{equation}\label{eq:omega_as}
\omega_{AS}=\omega_0\sqrt{\frac{1-\beta}{1-\alpha}}.
\end{equation}

The described resonance splitting is well known in the theory of
harmonic oscillators. Generally, bringing together two oscillators
of the same resonant frequency introduces coupling between them,
which results in splitting into two modes. Examples have been
shown of SRR resonant frequency as a function of some coupling
parameter, e.g.\ mutual orientation\cite{Hesmer2007} or twist
angle\cite{Liu2009}, and typically demonstrate a splitting or
hybridization of modes.

As we see, the direction and strength of the resonance shift are
determined by the coupling constants $\alpha$ and $\beta$. To
evaluate them, we use the expression for the
electromagnetic energy following from the Lagrangian \eqref{eq:L12}:
\begin{equation}\label{eq:E}
E=A(I_1^2+I_2^2+2\alpha I_1I_2)+B(Q_1^2+Q_2^2+2\beta Q_1Q_2).
\end{equation}
The first group of terms gives the magnetic energy and the second
group describes the electric energy.

A possible route to calculate $\alpha$ and $\beta$
is to approximate the electric and magnetic response of each ring
by a few terms of the multipole expansion.
The problem with this approach is that it is based on the
assumption that the observer (i.e.\ the second SRR) is at a large
distance compared to the dimensions of the source.  This
requirement is strongly violated in our metamaterial samples,
where the separation between rings is actually much smaller than
the outer ring diameter. This is essential for achieving strong
tuning by lattice manipulation.

Therefore we have chosen to calculate $\alpha$ and $\beta$
numerically from the known charge and current distributions,
$q(\mathbf{x})$ and $\mathbf{j}(\mathbf{x})$, of the fundamental
mode in a single SRR. Indeed, in the single mode approximation,
the energy of a pair of SRRs reads
\begin{multline}\label{eq:EW}
E = Q_1^2W_{E,11} + Q_2^2W_{E,22}+ 2Q_1Q_2W_{E,12} +\\
I_1^2W_{H,11} + I_2^2W_{H,22} + 2I_1I_2W_{H,12},
\end{multline}
where the parameters:
\begin{equation} \label{eq:WE}
W_{E,mn} = \int_{V_m}\!\!\!\mathrm{d}^3x\int_{V_n}\!\!\!\mathrm{d}^3x'\frac{q(\mathbf{x})q(\mathbf{
x'})}{4\pi\epsilon_0|\mathbf{x} - \mathbf{x'}|},
\end{equation}
\begin{equation} \label{eq:WH}
W_{H,mn} = \int_{V_m}\!\!\!\mathrm{d}^3x\int_{V_n}\!\!\!\mathrm{d}^3x'\frac{\mu_0\ \mathbf{j}(\mathbf{
x})\cdot \mathbf{j}(\mathbf{x}')}{4\pi|\mathbf{x} - \mathbf{x}'|}.
\end{equation}
The integrals can be easily evaluated once the charge and current
distributions are known. The integrations over $\mathbf{x}$ and
$\mathbf{x}'$ are over the same ring if $m=n$ or over different
rings otherwise. Accordingly, $V_1$ is a volume containing only
the first ring, and $V_2$ is a volume containing only the second.
The singular terms at $x=x'$ are handled using the analytical formulas
given in Ref.~\onlinecite{Arcioni1997}.

Comparing Eqs. \eqref{eq:E} and \eqref{eq:EW} shows that the coupling
parameters can be evaluated as
\begin{equation} \label{eq:coefficients_energy}
\alpha = \frac{W_{H,12}}{W_{H,11}}, \ \ \beta =
\frac{W_{E,12}}{W_{E,11}}.
\end{equation}

For comparison purposes, when inductive coupling is the dominant interaction mechanism
between the SRRs in a metamaterial, we are able to consider an array
of split rings as an array of current loops with some mutual
inductance between them. This approach can then be used to define the effective
permeability of a metamaterial sample\cite{Gorkunov2002}.
In particular, for thin wire loops with their axes oriented in the same
direction, the mutual inductance can be found\cite{Landau} by numerical integration:
\begin{multline*}
 L_{nn'}(\mathbf{r})=\frac{\mu_{0}r_{0}^{2}}{4\pi} \intop_{0}^{2\pi} \intop_{0}^{2\pi} \mathrm{d} \varphi_{1} \mathrm{d} \varphi_{2}
 \cos(\varphi_{1}\text{\textminus}\varphi_{2}) \times \\
 \left[\rho^{2}\text{+}z^{2}\text{+}2r_{0}^{2}(1\text{\textminus} \cos(\varphi_{1}\text{\textminus}\varphi_{2})) + 2\rho r_{0}(\cos\varphi_{2}\text{\textminus\ensuremath{\cos\varphi_{1}}})\right]^{-\frac{1}{2}},
\end{multline*}
where the distance vector $\mathbf{r}$ between the ring centers has
been decomposed into a radial component $\rho$ and axial component
$z$, $r_{0}$ is the ring radius and $\varphi_{1}$ and
$\varphi_{2}$ represent the angle about each ring.
We can use the calculated mutual inductance
to define an equivalent magnetic interaction parameter
\begin{equation} \label{eq:inductance}
\alpha_L = L_{nn'}/L
\end{equation}
(where $L$ denotes the self-inductance of one element)
which should approximate the interaction energy calculated from \eqref{eq:coefficients_energy}.
Asymptotically this interaction decays as $1/r$, so in a large
array the nearest neighbors provide the strongest contribution, but
do not necessarily dominate over all others. Clearly, this
interaction is highly anisotropic\cite{Sydoruk2006}, being positive
for rings on the same axis, but negative for rings in the same plane.

It is also possible to develop equivalent circuit models to calculate
the electric interaction between rings. For a pair of
coaxial rings, the total interaction can be reliably modeled as
a circular parallel conductor transmission line, as for
broadside-coupled split ring resonators \cite{Marques2003} or,
alternatively, with an extended circuit model accounting for the
distributed capacitance and inductance \cite{Shamonin2005}.
However, once we introduce some offset between the rings, these approaches
are not applicable and so we do not consider them here.

\section{Tuning interaction between a pair of split ring resonators\label{sec:pair}}

\begin{figure}[tb]
\includegraphics[width=1\columnwidth]{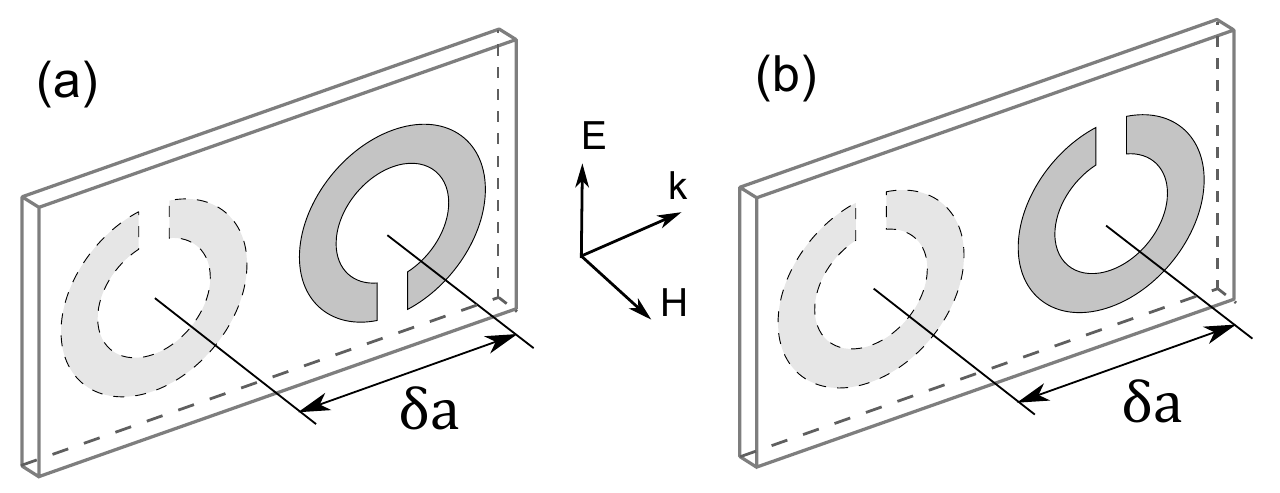}
\caption{Geometry of the pair of rings (a) broadside-coupled
and (b) gap-to-gap orientation\label{fig:2rings-geometry}}
\end{figure}

Having developed an approach for calculating near-field interaction
between a pair of rings, we now apply it to a canonical system
which has the basic properties of our experimental arrangement.
We consider a pair of SRRs, either identically (gap-to-gap) oriented or
rotated by 180\textdegree{} with respect to each other (as in the
broadside-coupled SRR), and subject to a lateral offset $\delta a$.
The geometry and incident polarization are shown in
Fig.~\ref{fig:2rings-geometry}. The rings have average radius $r_0 = 2.25$mm,
track width of 0.5mm, metal thickness of 0.03mm, gap width of 1mm,
and are separated in the transverse direction by 1.5mm. The resulting resonant frequency is 10.6GHz

\begin{figure}
\includegraphics[width=\columnwidth]{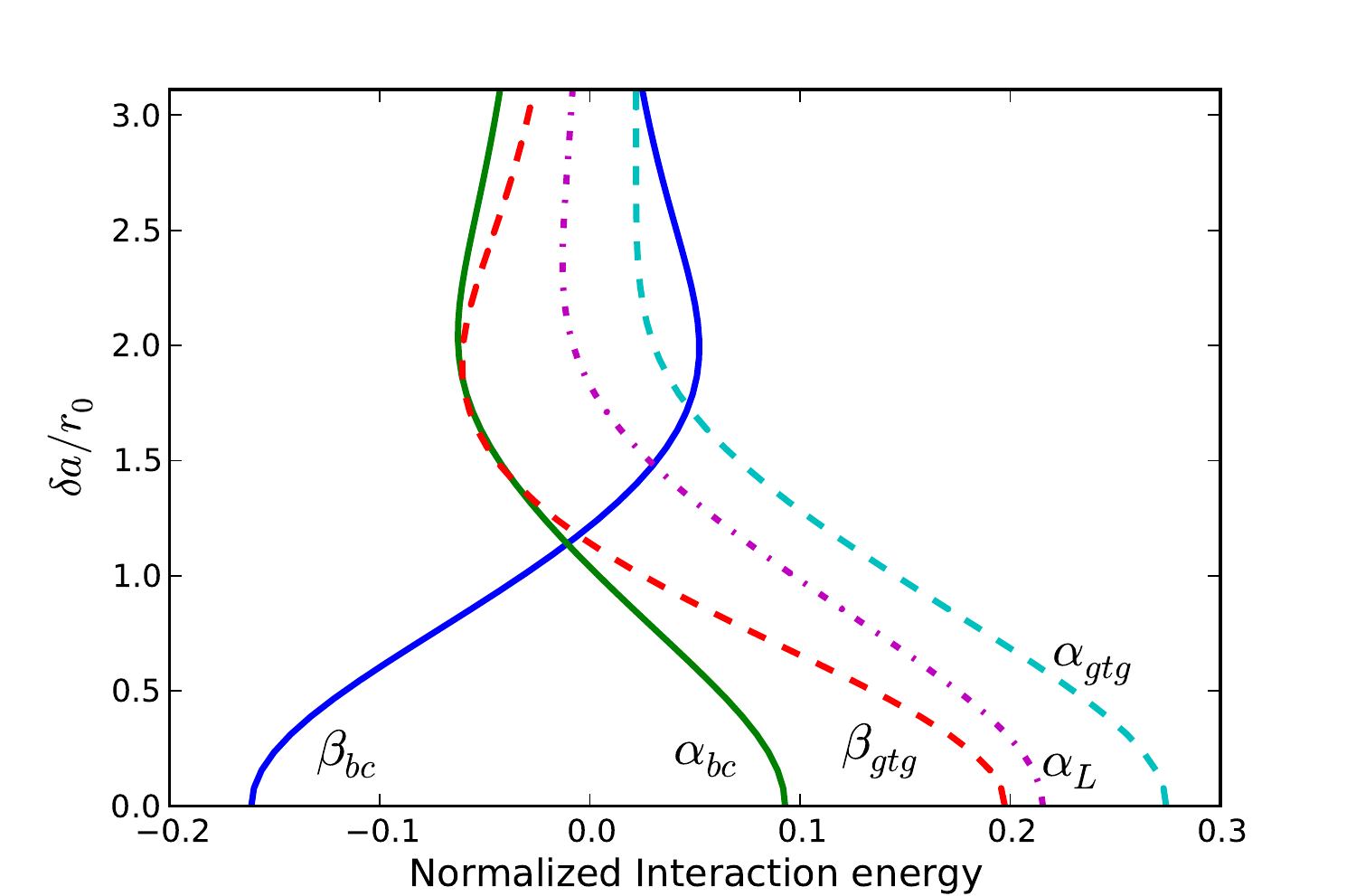}
\caption{(Color online) Magnetic ($\alpha$) and electric ($\beta$) coupling parameters for broadside coupled ($bc$)
and gap-to-gap ($gtg$) orientation of a pair of shifted SRRs. The magnetic interaction calculated from
the mutual inductance is given as $\alpha_L$.
\label{fig:interaction_energy}}
\end{figure}

We plot the interaction energy calculated from Eq.~\eqref{eq:coefficients_energy} in Fig.~\ref{fig:interaction_energy},
for different offsets between the rings.
It can be seen that the electric coupling parameter $\beta$ is nearly
symmetric between the two configurations. This can be understood from Fig. \ref{fig:current_charge},
where we see that the charge distribution has a strong dipole component oriented
in the $x$ direction. For the symmetric mode of the broadside-coupled orientation the charges accumulated on the
closest sides of the SRRs have opposite signs, the total charge distribution has the nature of
a pair of anti-parallel dipoles, thus $W_{E,12}<0$, and $\beta$ is also negative.
In contrast, in the gap-to-gap orientation, the closest charges are of the same sign,
the total charge distribution becomes like a pair of parallel dipoles and the
parameter $\beta>0$.

At an offset of about one ring radius ($\delta a \approx r_0$), the charge on one ring is
approximately equidistant from both positive and negative charges
on the opposite ring, thus the net coupling
passes through zero. At larger offsets, the electric coupling
changes its sign but remains smaller, since only the nearest halves of the SRRs
are effectively interacting, with this interaction decaying towards zero as the offset increases.

The magnetic interaction energy is also quite different for
the two orientations, with the magnetic interaction $\alpha_L$ calculated
by Eq. \eqref{eq:inductance} lying between the gap-to-gap and broadside curves. For the broadside-coupled case, the
situation is qualitatively similar to the analytical result $\alpha_L$.
At low offset, the magnetic field of one ring cuts through
the other ring in the same direction to the surface normal, thus
reinforcing the magnetic field and increasing the total energy. As
the offset is increased, the situation gradually shifts to become like a pair of
loops in the same plane, where the field from one ring cuts
through the other in the opposite direction with respect to the
surface normal. Hence, $\alpha_{bc}$ undergoes a change in sign.
However, in comparison to the ring with uniform current, the coupling
is substantially more negative. This is due to the current maxima
being on opposite sides of the rings, and hence further away from each other.

For the gap-to-gap orientation the magnetic interaction is much
stronger for low $\delta a$.  This is due to the current maxima
being located near each other which produces a
stronger contribution to the integral in $W_{H,12}$ thus
increasing $\alpha$. As the rings are further separated from each
other, the interaction energy reduces, but \emph{does not undergo
a change of sign}. We can intuitively understand this by
neglecting the small contributions of the current in the region
near the gaps, thus we effectively have two linear current
elements in the same plane which always interact with the same
sign. However, this balance is not universal and is determined by
the specific geometry and parameters. To check this, we studied a
geometry with a very small gap, so that the current distribution
was much more homogeneous, with lower resonance frequency. In this
simulation (not shown) the magnetic coupling did change sign and
both values of $\alpha$ converged closely to $\alpha_L$.

\begin{figure}[tb]
\includegraphics[width=1\columnwidth]{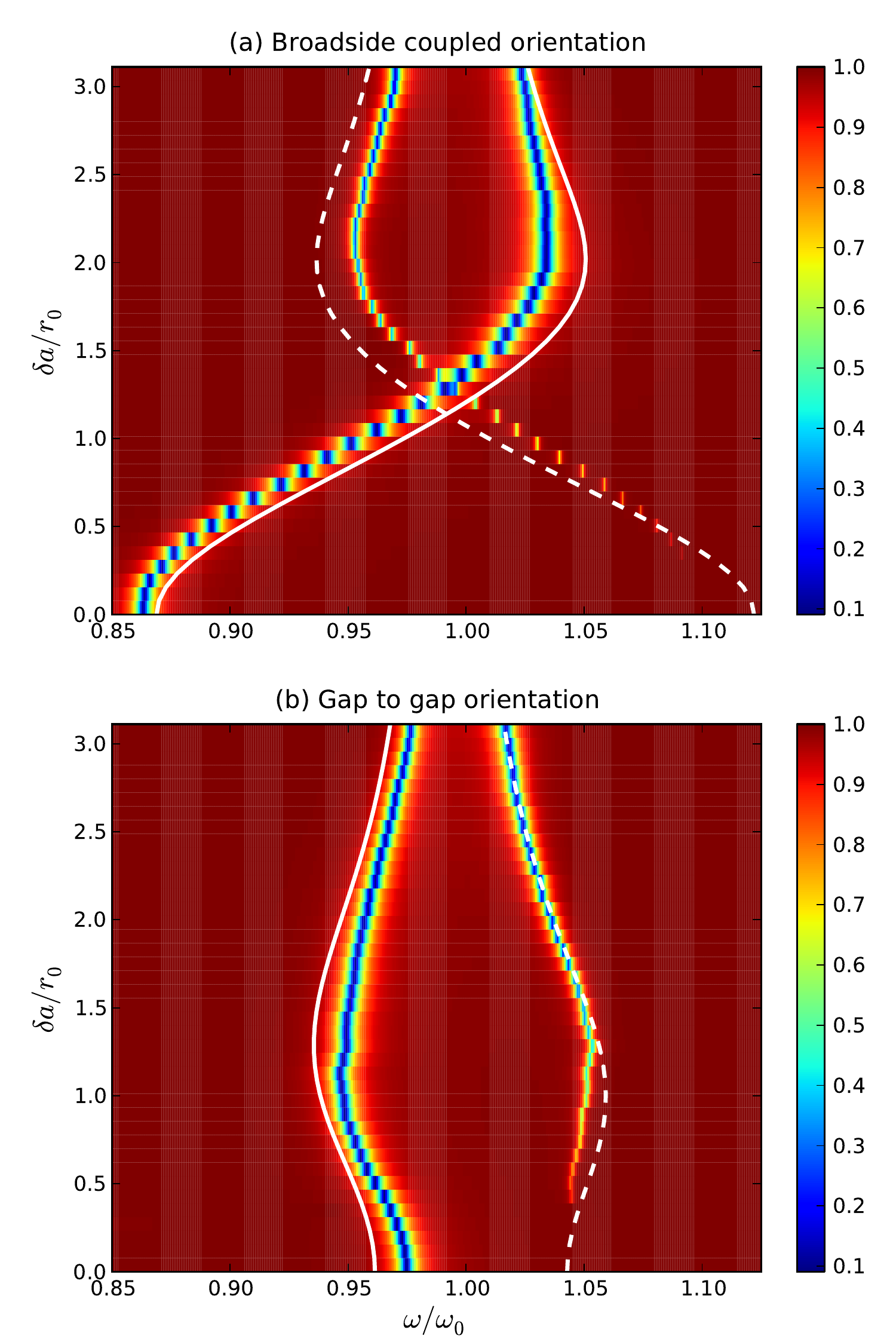}
\caption{(Color online) Numerical results.  Transmission
spectrum for a pair of (a) broadside-coupled and (b) gap-to-gap oriented rings.
Solid line: $\omega_{S}$ from Eq.~\eqref{eq:omega_s}, broken line: $\omega_{AS}$ from Eq.~\eqref{eq:omega_as}
\label{fig:freq-2rings}}
\end{figure}

In order to verify that the calculated coupling correctly
describes the frequency splitting of this system, we compare the
frequency shift predicted by Eq.~\eqref{eq:omega_s} and
\eqref{eq:omega_as} with that obtained from the full numeric
simulations. 
For consistency with our
interaction energy approach we assume a homogeneous free-space background.
We use the frequency domain solver of the
commercial software package CST Microwave Studio\cite{CST} to
model a pair of rings, in a unit cell with periodic boundary
conditions in the directions transverse to the propagation
direction. This periodic system enables us to define a transmission coefficient,
and the boundaries are 10mm from the rings. This value is chosen 
to be large enough so that there is no significant interaction 
with periodic neighbors, yet small enough to avoid significant scattering
into higher order diffraction modes. Thus we can consider this the limiting
case of a highly dilute metamaterial slab.

The transmission spectrum as a function of offset is plotted in
Fig.~\ref{fig:freq-2rings}. We see that all the important features
of the mode splitting are represented correctly by our single-mode
theory of coupled SRRs. Expansion of Eqs.~\eqref{eq:omega_s} and
\eqref{eq:omega_as} for small coupling predicts frequency splitting of
$\Delta \omega = \pm(\beta-\alpha)/2$, hence the curves are approximately
symmetric about $\omega_0$.

The strong splitting observed for the broadside-coupled orientation
is due to the opposite signs of the electric and magnetic coupling.
At $\delta a/r_0 \approx 1.1$ both $\alpha_{bc}$ and $\beta_{bc}$ change signs,
hence the crossing of the symmetric and anti-symmetric modes is observed.
In contrast, we see that for the gap-to-gap orientation, for small offset
$\alpha_{gtg}$ and $\beta_{gtg}$ are of the same sign, and thus they have an
opposing effect, resulting in small frequency splitting. Since
$\beta$ decreases much faster and changes sign, the result in maximum
frequency splitting for $\delta a/r_0$ between 1.1 and 1.5.
We note that in
Ref.~\onlinecite{Lapine2009} the resonant frequency based on
$\alpha_L$ was compared with experimental results for the gap-to-gap
orientation, and strong disagreement was found.

It can be seen that the calculated transmission through the cell
exhibits different depths of the resonance for the symmetric
and antisymmetric modes. This is due to the different efficiency of
coupling between the modes and the incident plane wave. For
instance, it is impossible to excite the anti-symmetric mode
with a normally-incident plane wave for $\delta a = 0$, since both rings
are excited in-phase. As the offset is increased, some retardation between
the rings occurs, and excitation of the anti-symmetric mode is allowed.

There are several reasons for the small quantitative disagreement
between the exact  calculations and those based on the calculated
interaction energy. Firstly, the minimum of
transmission occurs at a frequency slightly different from the
resonant frequency, due to coupling effects (impedance matching) between the incident
wave and the ring. Secondly,
there may be some small contribution of higher SRR eigenmodes due
to perturbation of the charge and current distributions. Thirdly, there may still be
some small influence of the periodic boundaries. Finally, our developed relations neglect retardation,
which is strictly valid in the sub-wavelength limit, whereas the
the outer radius of the rings is $0.18\lambda$ at $\omega_0$.
Retardation has previously been shown to modify the interaction
between SRRs through its influence on the dispersion of magneto-inductive
waves in arrays\cite{Radkovskaya2006,Zhuromskyy2009,Decker2009b}.

We emphasize that our approach developed in \prettyref{sec:theory}
is advantageous over the direct numerical calculation. Firstly, once
the mode profile is known, calculation of the frequencies in Fig. \ref{fig:freq-2rings}
takes approximately 30 seconds on a single CPU, whereas the direct
calculation of the full spectrum takes several hours on a multi-core
machine. Secondly, we are clearly able to demonstrate the nature of
the coupling, which yields insight into the tuning behavior.

\section{Tuning interaction in a bulk metamaterial \label{sec:slab}}

\begin{figure}[tb]
\includegraphics[width=1\columnwidth]{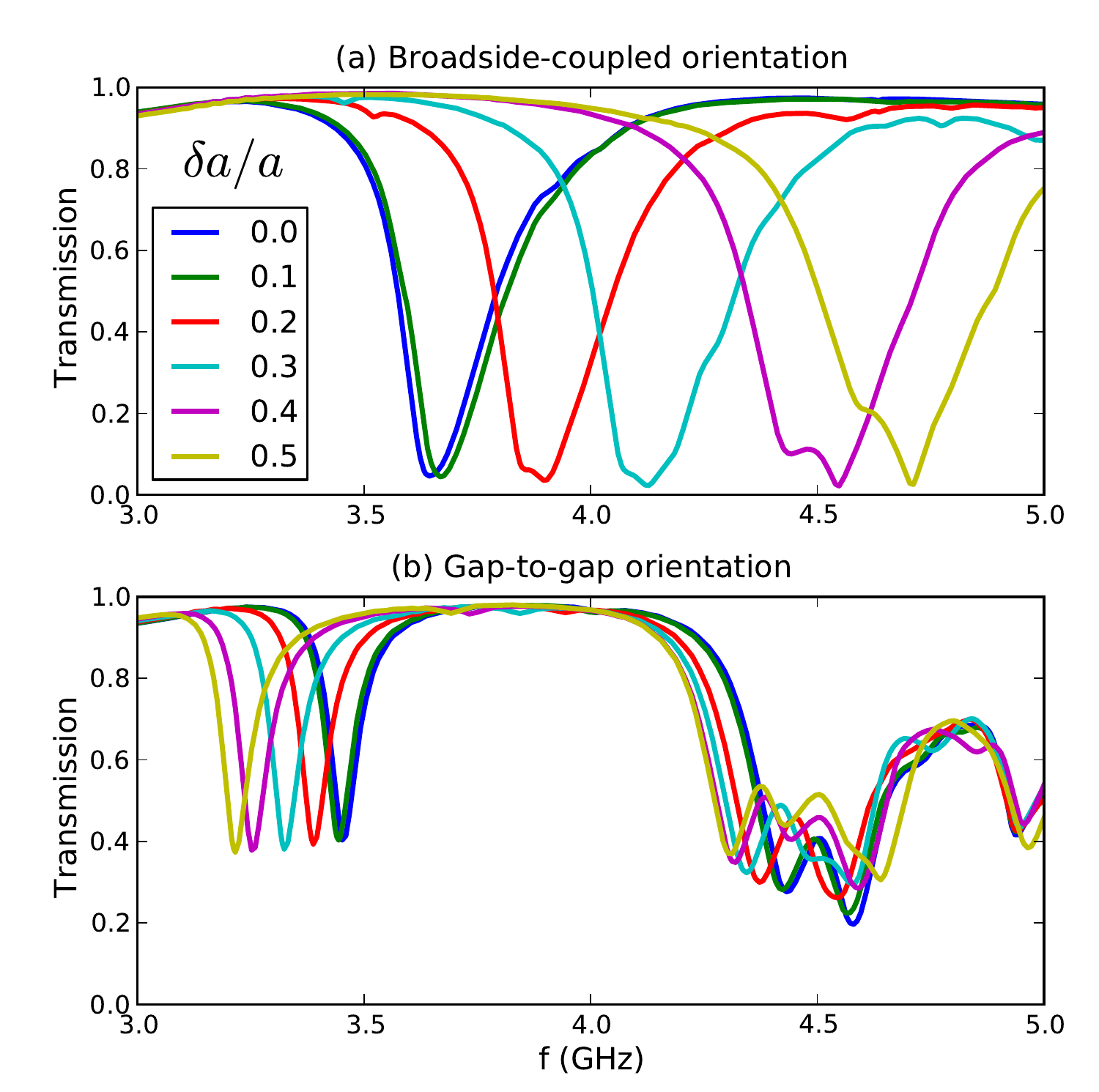}
\caption{(Color online) Experimental transmission while tuning $\delta a$ for split ring
resonator slab in waveguide, for (a) broadside-coupled and (b) gap-to-gap
orientation of adjacent layers.
\label{fig:experimental-tuning}}
\end{figure}

\begin{figure}[tb]
\includegraphics[width=1\columnwidth]{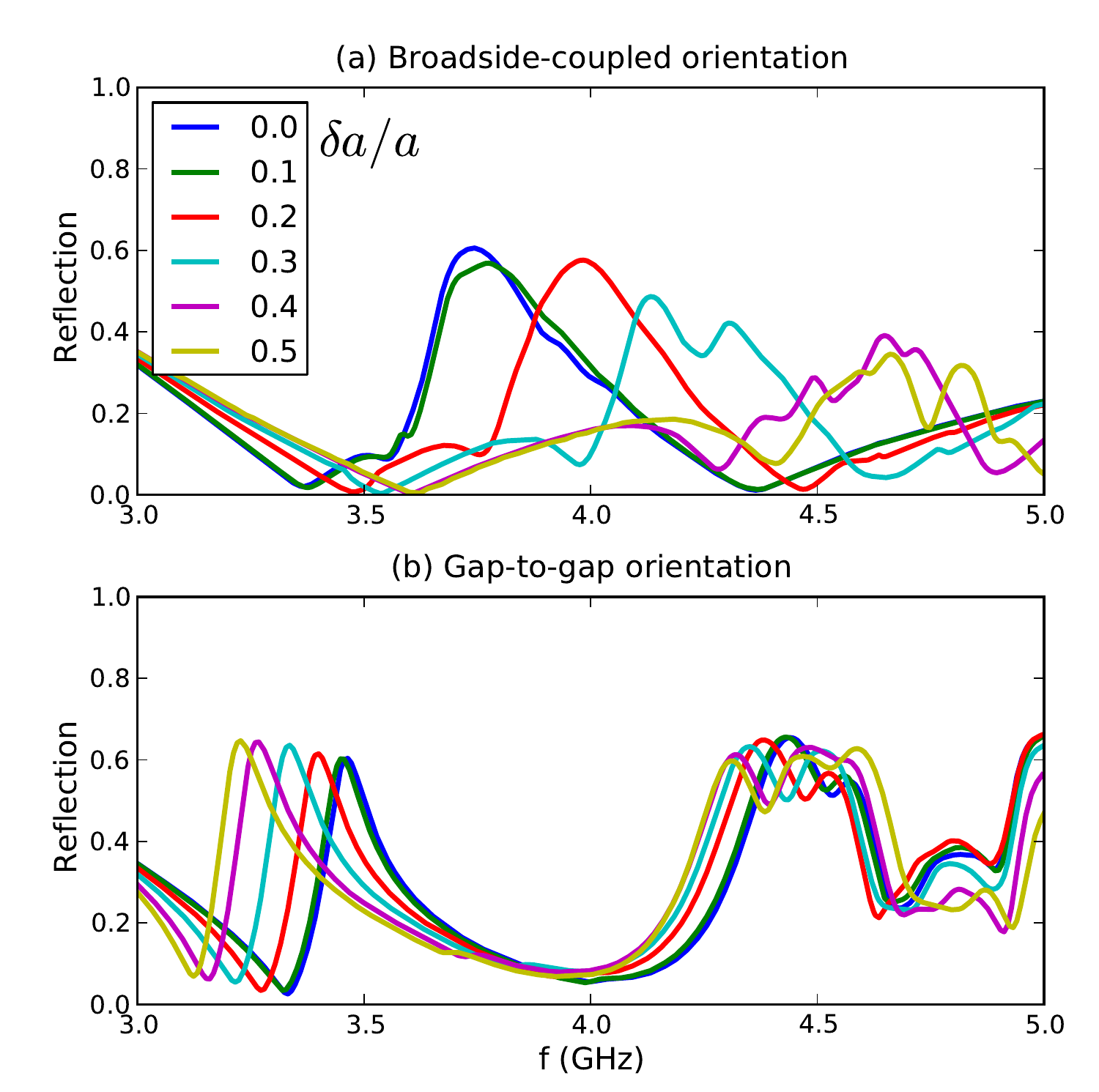}
\caption{(Color online) Experimental reflection while tuning $\delta a$ for split ring
resonator slab in waveguide, for (a) broadside-coupled and (b) gap-to-gap
orientation of adjacent layers.
\label{fig:experimental-reflection}}
\end{figure}

We now apply our approach to explain experimental results for tuning the response of a slab
of metamaterial. The metamaterial is fabricated using photolithography
to etch copper tracks onto FR4 printed circuit board, using the same
geometry as in our numerical simulation of a pair of rings. The fabricated
sample has 30 layers, each with 5 rings in the propagation direction
and is only one ring in height (i.e.\ a $5\times30\times1$ array).
The longitudinal period of the sample is 7mm, the transverse period is
dictated by the sample thickness (1.5mm) as there is no spacing between boards.
As with the pair of rings, we have assembled slabs with two relative orientations
of the split rings in adjacent planes --- gap-to-gap and broadside-coupled.
The sample is placed in the center of a WR229 rectangular metallic waveguide, with
dimensions 58.17$\times$29.08mm, excited at its dominant TE$_{10}$ mode.
We removed the influence of the coaxial adapters and feeding waveguide sections
by performing a TRL (through-reflect-line) calibration \cite{Engen1979}.
In Fig.~\ref{fig:experimental-tuning}
we show the experimentally obtained transmission through each slab of
closely coupled SRRs, with the corresponding reflection shown in
Fig.~\ref{fig:experimental-reflection}.
\begin{figure}[tb]
\includegraphics[width=1\columnwidth]{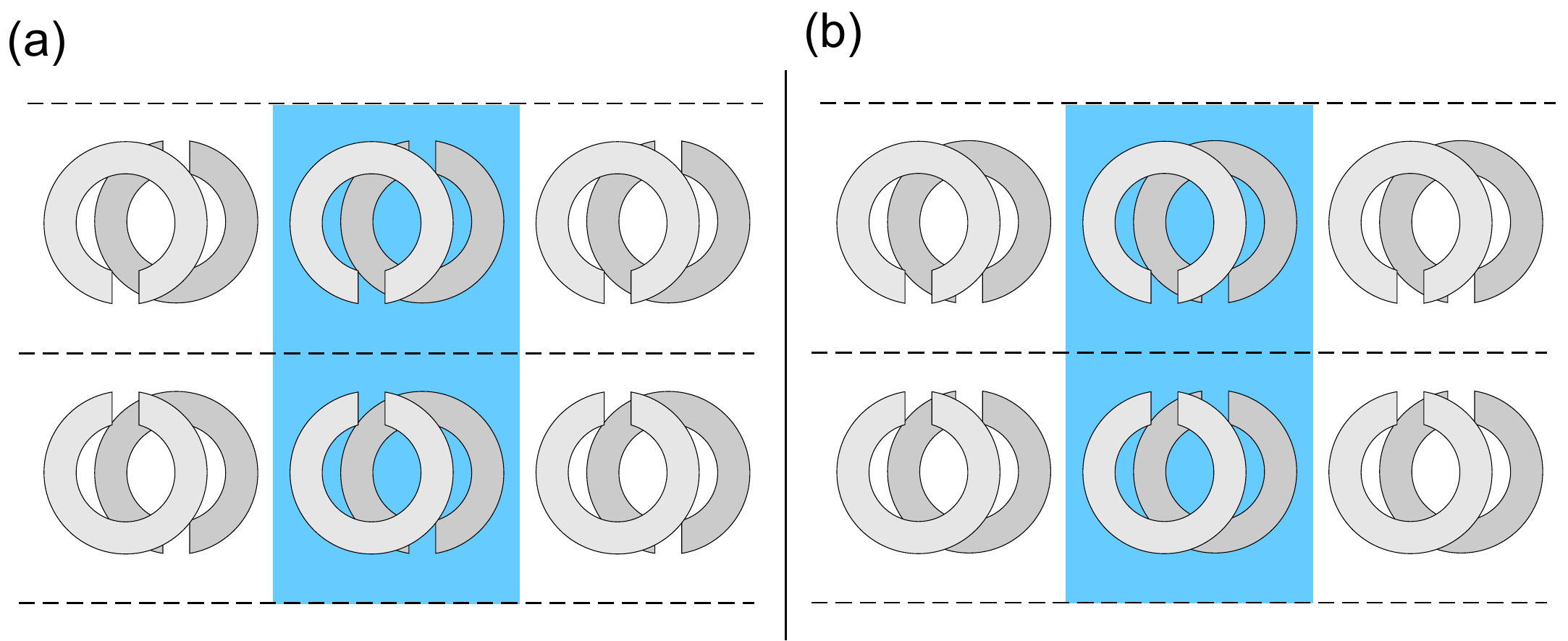}
\caption{Schematic of the effective super-lattice geometry corresponding to
the waveguide measurement for (a) broadside-coupled and (b) gap-to-gap
orientation. Dashed lines show planes of reflection
symmetry, and the shaded region shows the super-cell with
4 SRRs.\label{fig:superlattice-geometry}}
\end{figure}

\begin{figure}[tb]
\includegraphics[width=1\columnwidth]{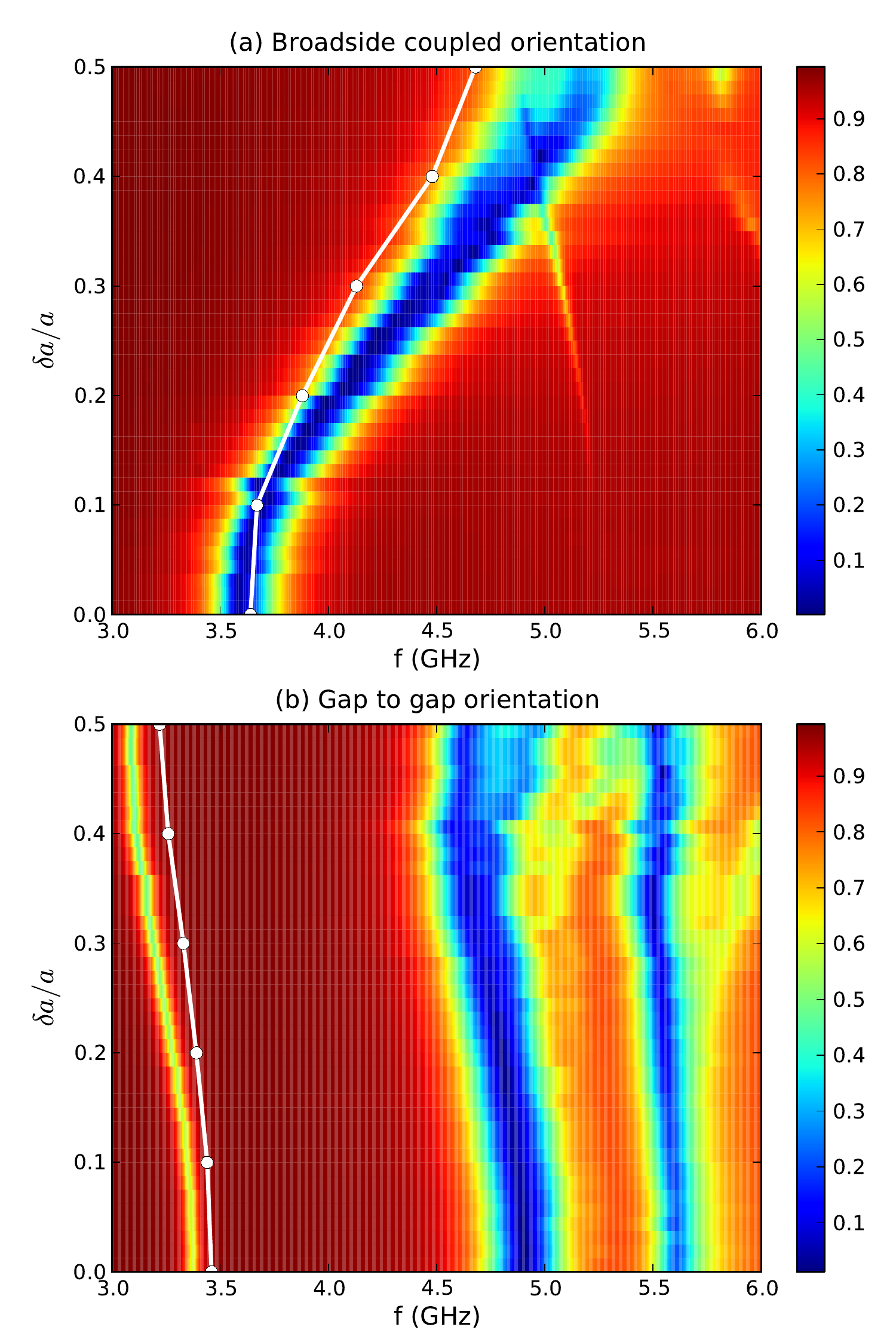}\caption{(Color online)
Numerically obtained transmission spectrum of metamaterial in waveguide,
(a) broadside-coupled and (b) gap-to-gap orientation. The white line indicates experimentally obtained resonant frequencies.
\label{fig:freq-waveguide}}
\end{figure}

The shift of resonant frequency shows good qualitative
agreement with the results for a pair of rings presented in \prettyref{sec:pair},
with very similar changes in the spectrum observed (noting that $\delta a /a = 0.5$
corresponds to $\delta a/r_0 = 1.56$).
However, for the gap-to-gap orientation, numerical simulation of a system of two boards
with 5 rings each, and \emph{periodic boundaries} in both transverse
directions (not shown), is qualitatively similar to the experimentally
observed results but quantitatively highly inaccurate.
The reason turns out to be the loss of symmetry when the system is
placed inside the waveguide, because
the upper and lower waveguide walls do not correspond to
periodic boundaries, but instead represent planes of mirror reflection.
Therefore, this system must be described as having a super-lattice arrangement in the vertical
as well as horizontal planes, with each super-cell consisting of four SRRs.
This cell has alternating orientation of the SRRs in the vertical direction
corresponding to the planes of mirror symmetry,
as shown in Fig.~\ref{fig:superlattice-geometry}(b).
Once this unit cell is taken into account, numerical simulations are in a good agreement with
the experiment (Fig.~\ref{fig:freq-waveguide}(b)).

Naturally, numerical simulations for the broadside-coupled orientation also
agree well with the experiment (Fig.~\ref{fig:freq-waveguide}(a)).
In this case a very similar result is provided with simple
periodic boundary conditions (not shown). For this orientation the super-lattice effectively formed by
the waveguide shown in Fig.~\ref{fig:superlattice-geometry}(a)
does not have an essentially different symmetry to the original super-lattice.

We can conclude that in
both cases the dominant mode of the slab corresponds to the dominant
symmetric mode of a pair of rings, with a similar pattern
of resonant frequency \emph{vs.}\ offset occurring.
The weaker coupling to the modes for the gap-to-gap orientation is due
to the shape of the symmetric mode. Its magnetic field has a large component
parallel to that of the waveguide mode, however its electric field
is primarily longitudinal, in contrast to the transverse electric field
of the waveguide mode.

In Fig.~\ref{fig:freq-waveguide}(b), we see two higher order modes,
which most likely correspond to the higher order resonances observed in
the experimental results in Fig.~\ref{fig:experimental-tuning}(b). From
the numerical simulations we observe that the current distributions of
these modes are symmetric, thus they correspond to higher-order modes of
the metamaterial slab, and not to the anti-symmetric mode of a pair of rings.
In contrast, in Fig.~\ref{fig:freq-waveguide}(a) there is a weakly coupled
anti-symmetric mode, which we verified by inspection of the currents.
This mode also qualitatively agrees with the corresponding mode of the
pair of rings, with somewhat weaker coupling due to the increased mismatch
to the incident waveguide mode.
This mode may correspond to some of the smaller features observable in
Fig.~\ref{fig:experimental-tuning}(a), however due to the size of these
features this cannot be reliably determined.

We do not consider offsets greater
than $0.5a$, since in an infinite lattice only shifts between 0 and 0.5
are unique, while in a finite structure, larger shifts result
in very irregular boundaries. Note that in the simulations we have neglected
the effect of the mode profile of the rectangular waveguide, which
would correspond to an effective variation of the angle of incidence
of the plane wave as a function of frequency, which can result in
a different response due to the anisotropy and non-negligible
spatial dispersion of the medium\cite{Simovski2008mm}.

Clearly the coupling in the complete lattice is much more complicated than
in the simple two-ring system, as the interactions between a large number of rings
must be taken into account.  In principle it is possible to extend the analysis of
 \prettyref{sec:pair} to an arbitrary number of rings.  However the
qualitative agreement between the experimental results and the modelled pair of rings
suggests that the phenomenology developed for the two rings is
generally applicable and leads to correct predictions.

Clearly, the coupling in the metamaterial lattice is much more complicated
than in the simple two-ring system, as the interactions between a large
number of rings must be taken into account. The qualitative agreement
between the experimental results and the modelled pair of rings suggests
that the phenomenology developed in \prettyref{sec:pair} for the two rings can be
extended to an arbitrary number of rings.  

Although an accurate generalization of our modelling approach to a bulk system lies beyond the scope of this paper, it is clear that the resulting homogenized effective metamaterial  parameters will
exhibit similar tuning pattern due to the resonance shift. Note that the
introduction of the effective parameters is justified when the ratio of the
unit cell size to the incident wavelength is small. Therefore, when
considering modifications of the lattice which create a super-lattice
structure, the size of the super-cell should be smaller than the wavelength.
There are homogenisation approaches in the literature (e.g. Ref.~\onlinecite{Petschulat2008}) which
include unknown parameters for interaction between resonant elements and our
semi-analytical approach would make an ideal tool for evaluating these
constants.

\section{Conclusion\label{sec:concl}}

We have analyzed the near-field coupling within metamaterials, considering both 
the relative orientation and the offset between the centers of two neighboring resonators.
Using a pair of split ring resonators as a simple model, we have shown the coupling mechanisms
at work in our recently proposed tuning scheme, based on the direct calculation of the interaction
energy. We have confirmed that these mechanisms can predict qualitatively the performance of a realistic
metamaterial structure. This paves a road towards a reliable design and development of 
tunable metamaterials for various applications.

We note that the specific geometry of the split rings can have a very
significant influence on the qualitative nature of the coupling,
including cases which run counter to our intuitive understanding
of current loops interacting magnetically.

Finally, we point out that the approach developed here for modeling
near-field effects is particularly promising for metamaterials
scaled down to operate at optical frequencies. In the visible, the
paradigm of ideally conducting metal fails and the area of
applicability of circuit models is rather limited. In contrast,
the consideration in terms of excitation and interaction of
plasmonic standing waves will provide a clear physical picture.

\section*{Acknowledgments}

The authors are grateful to Dr.~Lukas Jelinek and Prof.~Ricardo Marqu{\'e}s
(University of Seville) for helpful discussions. 
This work was supported by the Australian Research Council.
M.L. acknowledges hospitality of Nonlinear Physics Center and
support of the Spanish Junta de Andalusia Project P06-TIC-01368.
M.G. acknowledges support from the Russian Academy of Sciences,
BPS Program \textquotedblleft{}Physics of new materials and
structures\textquotedblright{}.


\end{document}